# Evaluating the impact of code smell refactoring on the energy consumption of Android applications

Hina Anwar
Institute of Computer Science
University of Tartu
Tartu, Estonia
hina.anwar@ut.ee

Dietmar Pfahl
Institute of Computer Science
University of Tartu
Tartu, Estonia
dietmar.pfahl@ut.ee

Satish N. Srirama
Institute of Computer Science
University of Tartu
Tartu, Estonia
satish.srirama@ut.ee

*Abstract*— **Energy consumption of mobile apps is a domain which is receiving a lot of attention from researchers. Recent studies indicate that the energy consumption of mobile devices could be improved by improving the quality of mobile apps. Frequent refactoring is one way of achieving this goal. In this paper, we explore the performance and energy impact of several common code refactorings in Android apps. Experimental results indicate that some code smell refactorings positively impact the energy consumption of Android apps. Refactoring of the code smells 'Duplicated code' and 'Type checking' reduce energy consumption by up to 10.8%. Significant reduction in energy consumption, however, does not seem to be directly related to the increase or decrease of execution time. In addition, the energy impact over permutations of code smell refactorings in the selected Android apps was small. When analyzing the order in which refactorings were made across code smell types, it turned out that some permutations resulted in a reduction and some in an increase of energy consumption for the analyzed apps. More research needs to be done to investigate how factors like size and age of software apps, and experience and number of contributors to app development correlate with (a) the number and type of code smells found and (b) the impact of energy consumption and performance after refactoring.**

## I. INTRODUCTION

Improving energy consumption of portable devices is a research domain that has captured the attention of the research community for quite some time. Among portable devices, mobile phone is the most commonly and widely used. "In 2018, total mobile phone sales were almost 1.9 billion units. In 2019, smartphone sales are on pace to continue to grow, at 5 percent year over year" [1]. According to an online report published in 2010, global mobile usage accounts for approximately one quarter of a percent of global mobile $CO_2$ emissions [2]. These numbers indicate that as the mobile usage is predicted to grow so will the carbon foot print associated with it. The situation could be improved by reducing the energy consumption of mobile phones. One way of reducing the energy consumption of mobile devices is to improve the software quality of mobile apps. This could be achieved by constant refactoring. Refactoring is "the process of changing a software system in such a way that it does not alter the external behavior of the code yet improves its internal structure. Code smells are an indicator of a problem in software design and quality that requires refactoring" [3]. Refactoring code smells could impact the energy consumption of mobile apps. We chose open source apps in Android for this investigation because Android holds 87.8% of the world's smartphone market share according to a Gartner report [4]. We assume that the types of code smells identified by Fowler et al. [3] occur in Java code independent of platform, although the frequency with which they occur might be distributed differently [5]. Therefore, we picked the set of code smell types analyzed in our study, as well as their refactorings, from Fowler et al.'s list.

In this paper, we investigate the energy impact of code smell refactorings on native Android open source apps. Code smell refactoring in Android apps has been discussed with respect to quality characteristics such as performance and maintainability [6], [7] but the energy impact of code smells has not yet been fully explored [8]–[11]. Studies by Pinto and Kamei [12], Yamashita and Moonen [13] indicate that developers do care about code smells (such as 'Long Method', 'Feature Envy') and conduct refactoring. The research presented in this paper extends previous studies by investigating the energy impact of refactoring five code smell types (first individually per type and then in permutation) of native Android open source apps. We also studied the effect of code smell refactorings on execution time. This study is intended to add to the findings of Verdecchia et al. [14] by exploring the energy impact of code smell refactoring on three native Android apps. We expect that our findings will help developers improve their Android apps not only with regards to maintainability but also in terms of energy efficiency.

The rest of the paper is organized as follows. In Section II we present the related work. Section III provides details of the research method, hypotheses, and the experimental setup. Section IV presents the results. Section V presents a discussion of the results. Section VI presents threats to validity. Finally, we summarize our findings and propose future directions of research in Section VII.

## II. RELATED WORK

Several empirical studies have investigated the energy impact of code smells in the mobile app domain.

Sahin et al. [15] applied code refactoring on nine Java applications and concluded that code refactoring could increase or decrease an application's energy usage. This study did not focus on code smells but instead directly applied six code refactorings from Eclipse IDE refactoring tool to the source code and evaluated their energy consumption.

Morales et al. [16] analyzed eight types of anti-patterns ('Blob', 'Lazy Class', 'Long-parameter list', 'Refused Bequest', 'Speculative Generality', 'Binding Resources too early', 'HashMap usage', 'private getter/setters') on 20 open-source Android apps to analyze the impact of anti-patterns and anti-pattern types on energy consumption. They proposed a novel anti-pattern correction approach called EARMO and evaluated its performance and found for the one app they analyzed an extension of battery life by 29 minutes.

Rodriguez et al. [9] present an analysis to evaluate the trade-off between OO design and battery consumption of mobile apps using code smells that impact directly on object

This work is supported by the institutional research grant IUT20- 55 of the Estonian Research Council and the Estonian Center of Excellence in ICT research (EXCITE).

creation and message passing between objects such as 'God Class', 'Brain Method', 'No Encapsulated field', 'No Self-Encapsulated field'.

Verdecchia et al. [14] discussed the energy impact of five code smells ('Long Method', 'Feature Envy', 'Type Checking', 'God Class', and 'Duplicated code') refactoring's on three ORM-based Java web applications. Their results indicate that in one out of three applications refactoring each code smell significantly impacted the energy consumption of the application.

Reimann et al. [17] published a catalogue of 30 Android code smells different from Fowler et al.'s [3] list of code smells. They introduced a tool 'Refactory' for detecting and suggesting fixes for the smells listed in their catalogue. Although they discussed energy efficiency of code refactoring, it is mostly speculation based on the literature, and empirical evidence through experimentation is needed to validate those claims.

Castillo et al. [10] analyzed the energy impact of 'God Class' refactoring on two Java applications. Their results indicated that God class refactoring has a negative impact on energy consumption.

Tufano et al. [18] and Delchey et al. [19] discussed when and how code smells appear in source code during development. They analyzed the circumstances in which code smells are introduced, the severity of their effects and how developers deal with it.

Some empirical studies profiled Android apps and evaluated code smells energy impact using novel software based tools. Carette et al. [20] conducted an empirical study on five open-source Android apps using the proposed automated approach called HOT-PEPPER, which enables developers to detect and correct three Android-specific code smells ('Internal Getter/Setter', 'Member Ignoring Method', and 'HashMap Usage'). Their results indicate a 4.38% global reduction in one out of five selected apps when the selected code smells were refactored.

Di Nucci et al. [21], [22] introduced a novel software based tool PETrA, for measuring the energy consumption of Android apps and compared its accuracy with the hardware toolkit by profiling 54 mobile apps from a public dataset. The results indicated that their tool performed equally to the hardware toolkit despite using any hardware component.

In this paper, we investigate the energy impact of code smell refactoring for five common code smells in Android apps. The work closest to this paper is the work published by Verdecchia et al. [14] in which they measured the energy consumption of five code smell refactorings on web applications. Our work differs from previous studies in that we focus exclusively on native open source Android apps. In related studies, where code smells in Android apps were explored, the focus was on Android specific code smells related to specific mobile resources and their correlation with performance, maintainability, and sustainability. In this paper, we focus on Java related code smells and their energy impact, as the programming language used in Android apps is also Java.

We selected native Android apps for this investigations because Android has the biggest share in the smartphone market, i.e., 87.8% [4]. According to an online report [23], published in 2017, out of all commerce transactions in North America, 44% was done using mobile apps, which means native mobile apps were preferred by users and derived three times more sales than desktop and web applications. In a survey on stack overflow data, it was revealed that java was the second most popular programming language among the Android developers [24]. From the overview of the state-of-the-art presented in this section, the evidence regarding the energy impact of Java code smells in native Android apps is limited and merits further research. Based on the statistics presented in the above mentioned surveys we expect our findings to be interesting for Android developers.

III. RESEARCH METHOD

In this section, we introduce the research questions, identify the code smells for refactoring, present the tools used for the detection of code smells, and define the experimental setup of our study.

*A. Research Questions*

*RQ1: Is there an impact of code refactoring on the energy consumption of Android apps?*

$H1_0$: The energy consumption between the original version and all possible refactored versions of the app remains the same.

$H1_1$: The energy consumption between the original version and some or all possible refactored versions of the app is not the same.

RQ1 investigates the energy impact of code refactorings from two aspects 1) impact, i.e., how do code refactorings impact the energy consumption of Android apps and 2) Consistency, i.e., are the observed effects of code smell refactorings in Android apps similar to the effects observed in other Java based applications discussed in related work.

*RQ2: Is there an impact of code refactoring on the execution time of Android apps?*

$H2_0$: There is no significant impact on execution time in any app versions if code refactoring is applied.

$H2_1$: There is a significant impact on execution time in all or any one of the app version if code refactoring is applied.

RQ2 investigates the impact of code refactoring on the apps execution time, i.e., What is the correlation between performance and energy efficiency?

*B. Code Smells*

In our study we focused on some commonly occurring code smells in Java applications and investigate their energy impact in Android apps. The selected smells are 'Long Method', 'Feature Envy', 'Type Checking', 'Duplicated Code', and 'God Class'. These smells have previously been investigated for their energy impact in some studies (as discussed in section 2 of this paper), but the results from those studies are limited by the type and number of apps on which these code smells were analyzed and hence merit further research.

*a) Long Method*

The 'Long Method' code smell is one of the most commonly occurring smells in Android apps. As defined by Fowler [3], in this smell there is a method which has become too long over time and could affect the maintainability of the app. Such methods could also be very hard to understand if you are not familiar with the code. Several refactoring techniques are defined by Fowler to get rid of this smell. In this paper, we are applying the 'Extract Method' refactoring for this smell. As a result of this refactoring, the 'Long method' will be divided into smaller methods which are called inside the original method.

### b) Feature Envy

The 'Feature Envy' code smell occurs when one class envies the features of another class, i.e., a method in one class sends many 'get method' calls to another class, which indicates that this method could be placed inside the other class whose features it envies [3]. In this paper, we apply the 'Move method' refactoring for this code smell.

### c) Type Checking

The 'Type Checking' code smell occurs when an attribute in a class representing the state is checked for different values and based on its value branch of a switch or if-else statement is executed. This state attribute is referred to as a type field. The term 'Type Checking' was used by the creator of JDeodorant for this condition [3], [25]. In this paper, we have used the refactoring 'replace type code with state/strategy'. This refactoring results in new methods and subclasses, making it a candidate for our investigation.

### d) Duplicated Code

The 'Duplicated Code' code smell occurs when the same code is used in many places inside an app. It could be intentional or it could be due to the copying of the code or it could be the result of another refactoring. It could be refactored by combining the different code clone structures into one [3].

### e) God Class

The 'God Class' code smell occurs when over time many functionalities are added to the same class in a project, making it responsible for a large percentage of the app's architecture. This smell is removed by dividing the large class into smaller classes, each having a unique functionality [3].

## C. Code Smell Detection tool and Refactoring

We selected the code smell detection and refactoring tool based on the list of activities that should be supported by any good refactoring tool as proposed by Mens et al. [26]. Verloop [27] compared the quality of four code smell detection tools JDeodorant, UCDectctor, Checkstyle, and PMD. Fontana et al. [28] compared the performance of Eclipse, Intellij IDEA, JDeodorant, RefactorIT. Tsantalis et al. [29] explained how JDeodorant covers all the refactoring activities. Paiva et al. [30] compared four tools, i.e., inFusion, JDeodorant, PMD, and JSpIRIT to evaluate their accuracy and agreement. Based on these studies we chose to use JDeodorant for code smell detection. JDeodorant follows all the refactoring activities stated by Mens et al. According to the studies mentioned above JDeodorant is very effective in detecting 'Long Method', 'Feature Envy', 'God Class', 'Duplicated Code', and 'Type Checking'. We manually evaluated each of the refactorings suggested by JDeodorant before applying them to ensure that they do not change the external behavior of the app.

## D. Selected Apps

We selected and downloaded from F-Droid open source native Android apps that were more than two years old, had more than 10K downloads, and at least a rating of '4' in the Google play store. The selected apps were picked from the categories 'Tools' and 'Puzzles'. The detection of code smells and the generation of refactoring candidates was done using JDeodorant with Eclipse IDE. Table I summarizes the characteristics of each of the selected apps. The first row (Years) shows the age of the project (up till February 2019) for example, the age of 'Calculator' app is 6 years and 8 months. The second row (LOC) shows the source line of codes and gives an estimate of the size of the project. The third row (Contributors) shows the number of developers working on the project. The fourth row (Commits) shows the number of commits made to the project by contributors of the project. The fifth row (Downloads) shows the number of times the selected app has been downloaded by users from the Google play store. The last row (Ratings) shows the mean of ratings given to the app by Google play store users.

TABLE I. CHARACTERISTICS OF SELECTED ANDROID APPS

|              | Calculator [31] | Todo-List [32] | Openflood [33] |
|--------------|-----------------|----------------|----------------|
| Years        | 6.8             | 3              | 3.1            |
| Contributors | 18              | 7              | 7              |
| LOC          | 7758            | 6145           | 1236           |
| Commits      | 1142            | 258            | 138            |
| Downloads    | 1,000,000+      | 1000+          | 10,000+        |
| Ratings      | 4.5             | 4              | 4.6            |

## E. Testing Tool

We used Espresso as it comes built-in with Android Studio and, according to a recent study [34], is very fast and reliable outperforming other tools for testing native Android apps. Since we do not intend to navigate outside the app under test, Espresso is a good choice as it supports white box testing and UI tests can be created easily. The test scripts created in Espresso do not need time delays to function properly. Therefore, the overhead introduced in the execution of the apps when using Espresso is low.

We used our test scripts for energy measurements and also to ensure correctness of the refactored app code. The test cases[1] included in the test scripts may seem trivial but they were solely defined to ensure that the code containing a smell is actually executed. To validate that the test scripts triggered the execution of code containing a code smell as well as the execution of the corresponding refactored code, execution traces were inspected. The method defined by the Android debug class [35] was used to activate the generation of the execution trace.

## F. Energy Measurement

We used Monsoon power monitor to measure the energy consumption, recording the power measurements at a rate of 5KHz. The time between two readings was 0.0002 seconds. In this experiment, energy consumption corresponds to the total amount of energy used by the mobile device within a period of time. Energy is measured in Joules which is power (watts) times measurement period (seconds). We calculated the energy associated with each reading as E = Power x (0.0002). The total energy consumption corresponds to the sum of the energy associated with each reading.

---

[1] https://bitbucket.org/hinaanwar2003/calculator/wiki/Test%20Cases
https://bitbucket.org/hinaanwar2003/todolist/wiki/Test%20Cases
https://bitbucket.org/hinaanwar2003/open_flood_src/wiki/Test%20Cases

Before starting the experiment, a baseline was recorded to measure the energy consumption of the mobile device in an idle state, which was then subtracted from the actual energy readings during the experiment to filter out the energy used by the app under test. During the experiment, the screen brightness was set to minimum and only essential Android services were run on the phone.

The timestamps from the adb logs recording during energy measurement and the CSV files containing energy readings were matched to mark the start and end of an execution run. Each app version was executed 10 times to account for underlying variation in the mobile device. While the UI test was executed we used the Monsoon power monitor to measure the energy consumption of each version.

*G. Test Environment*

The test environment consisted of an HP Elite Book, the Monsoon power monitor, and an LG Spirit Y70 Phone having Android 5.0.1 as the operating system with 1GB of RAM, and a 2100mAh battery. The test was controlled from the HP Elite Book using a script that automated the process and runs each app version 10 times. This saved manual effort and ensured that no problems were created during the experiment due to human error. The mobile phone was connected to the host computer with a USB cable via Monsoon power monitor which disables the USB phone charging once the energy reading starts. Battery terminals of the mobile phone were connected to the main channel of Monsoon power monitor. The mobile device was connected to the Monsoon power monitor according to the instruction manual of the power monitor [36].

---

Algorithm: Steps done for all readings and for all apps

1. Setup power monitor using python script
2. Install the app in the phone using adb.
3. Clear logcat and previous log files on the phone using adb.
4. Start power monitor readings using a python script from PC.
5. Start batch file on the phone for following operations
    a. Run the app
    b. Start recording the adb logs
    c. Play test scenario.
    d. Stop recording adb logs and transfer results to a text file.
    e. Stop app
6. Stop recording energy readings using python script
7. Transfer energy readings to CSV file
8. Pull log files from the phone using adb.
9. Uninstall app using adb.

---

*H. Experimental Setup and Execution*

In this section, we describe the setup of our experiment to measure the energy consumption of code smell refactorings on Android apps. Since Android studio projects cannot be opened directly inside Eclipse like Java projects, the apps were first modified using an Eclipse plugin in Android studio generating a file structure that helps Eclipse IDE to open the project with Gradle and recognize the Java files. Build path dependencies for projects opened in Eclipse IDE were solved manually by the author. This step was necessary because JDeodorant is provided as a plugin in Eclipse IDE and works only with Java projects. Code smell refactorings were applied in two ways: 1) code smell refactorings per code smell type and 2) code smell refactorings for all code smell types (one type after the other) in various permutations. For single code smell type we first detected the smells using JDeodorant and then candidate refactorings suggested by JDeodorant were applied. For each suggested refactoring, the test script was executed to ensure that the refactoring could be applied without altering the functionality of the app. If the test script failed for a suggested refactoring, that refactoring was ignored. We applied refactorings for a single code smell type and made new versions of the refactored app. For assessing whether the order of refactoring has an effect, we also made versions in which all code smell types were refactored in various permutations grouped by type. As the total number of all possible permutations of five code smell types was 120, we chose a sample of permutations randomly using Fisher-Yates shuffle [37]. For versions of the app where refactorings of all code smell types were applied in various permutations, we ignored any new candidate refactorings identified as a result of applying refactorings to a previous code smell type. For most code smell types, more refactorings could be applied but we only applied and reported the refactorings for which the test script was successfully executed.

Tables II, III, and IV show the numbers of applied code smell refactorings for each app. The names of code smells in the tables are abbreviated as LM (Long Method), FE (Feature Envy), TC (Type Checking), DC (Duplicated Code), and GC (God Class). In the first column 'Refactoring', the row 'Single Refactoring' refers to the situation where only code refactorings of one code smell type were applied. The rest of the rows in the first column represent the situation where refactorings of all code smell types were applied one by one in various permutations.

For statistical analysis, normality and homogeneity of the energy data were checked before doing the analysis of variance (ANOVA) to test the hypotheses related to RQ1 and RQ2. For our analyses, α was set to 0.05. We calculated the effect size using Cliff's delta [38].

TABLE II.    NUMBER OF REFACTORINGS FOR CALCULATOR APP

| Refactoring | LM | FE | TC | DC | GC |
|---|---|---|---|---|---|
| Single Refactoring | 41 | 1 | 3 | 3 | 6 |
| LM-TC-GC-FE-DC | 41 | 1 | 2 | 2 | 5 |
| FE-LM-TC-DC-GC | 41 | 1 | 2 | 2 | 5 |
| TC-GC-LM-FE-DC | 39 | 1 | 3 | 2 | 5 |
| GC-LM-DC-TC-FE | 41 | 1 | 2 | 2 | 6 |
| FE-DC-TC-GC-LM | 40 | 1 | 3 | 3 | 5 |
| TC-LM-DC-FE-GC | 39 | 1 | 3 | 2 | 5 |
| LM-DC-TC-GC-FE | 41 | 1 | 2 | 2 | 5 |
| LM-TC-DC-FE-GC | 41 | 1 | 2 | 2 | 5 |
| DC-TC-FE-LM-GC | 40 | 1 | 3 | 3 | 5 |

TABLE III.    NUMBER OF REFACTORINGS FOR TODO-LIST APP

| Refactoring | LM | FE | TC | DC | GC |
|---|---|---|---|---|---|
| Single Refactoring | 45 | 3 | 3 | 4 | 11 |
| LM-TC-GC-FE-DC | 45 | 1 | 3 | 2 | 6 |
| FE-LM-TC-DC-GC | 44 | 3 | 3 | 2 | 6 |
| TC-GC-LM-FE-DC | 41 | 1 | 3 | 2 | 11 |
| GC-LM-DC-TC-FE | 43 | 1 | 3 | 2 | 11 |
| FE-DC-TC-GC-LM | 36 | 3 | 3 | 4 | 11 |
| TC-LM-DC-FE-GC | 45 | 1 | 3 | 2 | 6 |
| LM-DC-TC-GC-FE | 45 | 1 | 3 | 2 | 6 |
| LM-TC-DC-FE-GC | 45 | 1 | 3 | 2 | 6 |
| DC-TC-FE-LM-GC | 40 | 3 | 3 | 4 | 6 |

TABLE IV. NUMBER OF REFACTORINGS FOR OPENFLOOD APP

| Refactoring | LM | FE | TC | DC | GC |
|---|---|---|---|---|---|
| Single Refactoring | 7 | 1 | 0 | 0 | 1 |
| LM-GC-FE | 7 | 0 | 0 | 0 | 1 |
| FE-LM-GC | 7 | 1 | 0 | 0 | 1 |
| GC-LM-FE | 6 | 0 | 0 | 0 | 1 |
| FE-GC-LM | 6 | 1 | 0 | 0 | 1 |
| LM-FE-GC | 7 | 1 | 0 | 0 | 1 |
| GC-FE-LM | 6 | 0 | 0 | 0 | 1 |

## IV. RESULTS AND EVALUATION

In this section, we present the results of our investigation regarding the impact of code refactoring on energy consumption and execution time for Android apps.

*RQ1: Is there an impact of code refactoring on the energy consumption of Android apps?*

An analysis of variance (ANOVA) showed that the effect of code smell refactorings on energy consumption of Android apps was significant in two out of three apps.

The hypothesis $H1_0$, stating that the energy consumption between the original version and all possible refactored versions of the app remains the same, was rejected for the 'Calculator' and 'Todo-List' apps. Table V shows the calculated p-values and F-values for each app. For the 'Calculator' and 'Todo-List' apps, the p-values are less than $\alpha=0.05$ and corresponding F-values are large indicating that the change in energy consumption after applying code smell refactorings was significant. For the 'Openflood' app the alternate hypothesis $H1_1$ was rejected as no significant difference in energy consumption was detected applying ANOVA, i.e., the p-value is not less than $\alpha$ and F-value is very close to 1.0.

TABLE V. ANOVA RESULTS FOR ALL APPS

| Application | p-Value | F-value |
|---|---|---|
| Calculator | 0.00005343 | 7.3503 |
| Todo-List | 0.0009859 | 4.3473 |
| Openflood | 0.4351 | 0.9666 |

Figures 1 to 3 show the boxplots of Treatment vs. Energy consumption in joules for each of the apps. The treatments are along the x-axis while energy consumption is shown along the y-axis. Treatments refer to the type of code smell refactoring applied (or=Original, LM=Long Method, FE=Feature Envy, TC= Type Checking, GC=God Class, DC=Duplicated Code, ALL=Avg. of all versions where the refactorings of all code smell types were applied in permutations). From figures 1 to 3 we see that for permutations of code smell refactorings, for all apps, there was no significant difference in energy consumption. For individual code smell refactorings, two out of three apps had a significant difference in energy consumption, but the direction of the effect is not uniform across treatments. Only TC, DC, FE, and LM consistently saving energy in both apps, however, the strength of the effect is not uniform (more details in the discussion section).

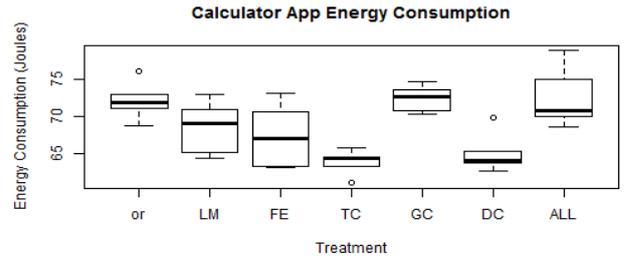

Figure 1: Energy consumption in joules for Calculator app per treatment

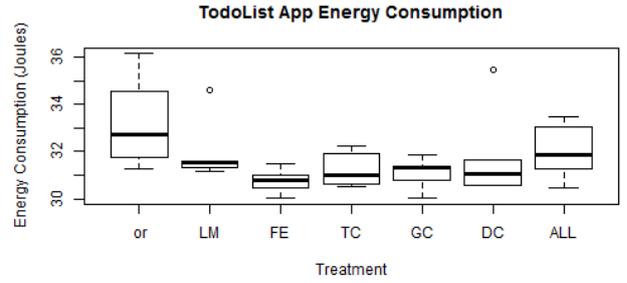

Figure 2: Energy consumption in joules for Todo-List app per treatment

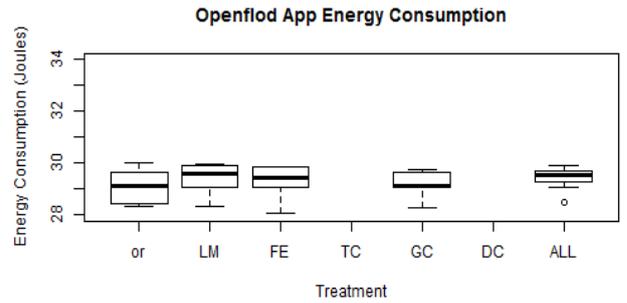

Figure 3: Energy consumption in joules for 'Openflood' app per treatment

*RQ2: Is there an impact of code refactoring on the execution time of Android apps??*

Based on our measurement we could not find a significant impact of code refactoring on execution time for the selected Android apps. The p-values were 0.2584, 0.113, and 0.384 for the 'Calculator', 'Todo-List', and 'Openflood' apps respectively, which was greater than the alpha value of 0.05. Therefore $H2_0$ stating that the execution time between the original version and all possible refactored versions of the app remains the same could not be rejected.

## V. DISCUSSION

Form the results related to RQ1 we see that there was a significant impact of refactoring code smells in two out of three Android apps. In the following, we take a closer look at how different configurations of code smell refactorings affected the energy consumption of each of the analyzed Android apps.

TABLE VI. OVERVIEW OF ENERGY CONSUMPTION RESULTS

| Application (joules) | | Treatments | | | | | | |
|---|---|---|---|---|---|---|---|---|
| | | Or | LM | FE | TC | GC | DC | ALL |
| Calculator | Median | 71.9 | 69.1 | 67.1 | 64.3 | 72.6 | 64.1 | 71.6 |
| | SD | 2.6 | 3.6 | 4.4 | 1.7 | 1.8 | 2.8 | 3.4 |
| | % | - | -3.8 | -6.6 | -10.5 | +1.0 | -10.8 | -1.3 |
| | ES | - | L | L | L | N | L | S |
| Todo-List | Median | 32.7 | 31.5 | 30.8 | 31.0 | 31.3 | 31.0 | 31.8 |
| | SD | 2.3 | 1.4 | 0.5 | 0.7 | 0.7 | 2.0 | 1.0 |
| | % | - | -3.6 | -5.9 | -5.2 | -4.4 | -5.1 | -2.6 |
| | ES | - | M | L | L | L | L | M |
| Openflood | Median | 29.1 | 29.5 | 29.4 | N/A | 29.1 | N/A | 29.5 |
| | SD | 0.6 | 0.5 | 0.5 | - | 0.4 | - | 0.4 |
| | % | - | +1.5 | +1.1 | N/A | -0.02 | N/A | +1.3 |
| | ES | - | S | S | N/A | N | N/A | S |

(SD=standard deviation, %= percentage change, ES=effect size, L=large, M= medium, S=small, N=negligible)

Table VI gives details about the median, standard deviation, percentage change, and effect size for each app after each treatment (we used the same abbreviation as in figures 1 to 3). We see that for the 'Calculator' app maximum reduction in energy consumption was recorded in app versions where 'Duplicated code' (10.8%) and 'Type checking' (10.5%) code smell refactorings were applied. Verdecchia et al. [14] reported maximum energy reduction in versions where 'Long Method' and 'Feature Envy' code smell refactorings were applied. In our experiment, the number of applied refactoring of both 'Type Checking' and 'Duplicated Code' code smells were smaller as compared to 'Long Method' code smell refactorings but the recorded effect size for energy reduction was larger. By running test scripts, we made sure that all refactored code was executed at least once. In the case of 'Long Method' some methods were called more often than the others, hence the higher frequencies of method calls and parameter passing might have caused medium energy reduction. In the case of the 'Openflood' app, we observed that 'Long Method' code smell refactoring increased the energy consumption but the effect size was too small to be considered significant. Verdecchia et al. [14] also reported an increase in energy consumption in two out of three projects when 'Long Method' code refactoring was applied. They too reported that the effect size was too small to be considered significant in those projects. The number of code refactoring for 'Long Method' code smell were also small in those two projects. However, one project, where 'Long Method' code refactoring significantly reduced energy consumption, the number of applied code refactoring was also considerably higher. We observed the same trend in our results.

> Insight 1: Impact of refactoring only a single type of code smell on energy consumption of selected apps was not consistent. However, in two out of three selected apps, where the effect size was medium or large, the energy consumption decreased due to refactoring.

Table VII shows the median, standard deviation, percentage change, and effect size when the refactorings of all code smell types were applied in permutations. We only analysed the 'Calculator' and 'Todo-List' app because for 'Openflood' app the change in energy consumption was not significant. In the 'Calculator' and 'Todo-List' apps. In 'Calculator' app the permutation 'LM-TC-DC-FE-GC' resulted in a maximum increase of energy consumption up to 7.25%, while the permutation 'DC-TC-FE-LM-GC' resulted in a maximum decrease of energy consumption up to 9.3%. In the 'Todo-List' app no permutation resulted in a significant increase in energy consumption, However, the maximum decrease of energy consumption was in permutation 'FE-DC-TC-GC-LM' (up to 7%).

TABLE VII. ENGERGY CONSUMPTION RESULTS WHEN A PERMUTATION OF CODE SMELL REFACTORINGS WAS APPLIED.

| Refactoring | Calculator | | | | Todo-List | | | |
|---|---|---|---|---|---|---|---|---|
| | Med | SD | % | ES | Med | SD | % | ES |
| Original (baseline) | 71.9 | 2.6 | - | - | 32.7 | 2.0 | - | - |
| LM-TC-GC-FE-DC | 74.8 | 6.3 | +4.0 | N | 31.3 | 0.9 | -4.3 | L |
| FE-LM-TC-DC-GC | 66.8 | 2.8 | -7.0 | L | 31.4 | 0.8 | -3.9 | M |
| TC-GC-LM-FE-DC | 66.8 | 6.9 | -7.1 | S | 31.3 | 0.9 | -4.1 | L |
| GC-LM-DC-TC-FE | 75.8 | 6.9 | +5.4 | S | 32.3 | 2.3 | -1.1 | N |
| FE-DC-TC-GC-LM | 76.6 | 6.5 | +6.9 | L | 30.4 | 0.5 | -7.0 | L |
| TC-LM-DC-FE-GC | 70.9 | 2.5 | -1.3 | S | 33.8 | 0.7 | +3.4 | N |
| LM-DC-TC-GC-FE | 69.4 | 4.8 | -3.4 | S | 32.2 | 1.5 | -1.3 | N |
| LM-TC-DC-FE-GC | 77.1 | 4.8 | +7.2 | L | 30.9 | 0.5 | -5.3 | L |
| DC-TC-FE-LM-GC | 65.1 | 9.6 | -9.3 | L | 31.9 | 1.2 | -2.4 | S |

(Med=median, SD= standard deviation, %= percentage change, ES=effect size, L=large, M= medium, S=small, N=negligible)

To explore the possible reasons behind the difference in energy consumption between permutations, we additionally calculated some of the C&K metrics [39], i.e., LOC, number of methods (NM), number of classes (NC), cyclomatic complexity (CC), and coupling between objects (CBO) for the 'Calculator' and 'Todo-List' apps using the Metrics plugin in Eclipse IDE. We observed that compared to the original versions of the apps the LOC, NM, NC, and CBO measures increased in all versions of the 'Calculator' and 'Todo-List' apps where the refactorings of all code smell types were applied in permutations. However, CC decreased in all these versions. The values for code metrics were different among different versions and apps indicating that the internal design and thus the resource usage during execution changed, causing a change in energy consumption. In our experiment, we also observed that some permutations like 'LM-TC-DC-FE-GC' and 'FE-DC-TC-GC-LM' impacted energy consumption differently in both apps (see Table VII). If not used carefully the refactorings for different code smells could cancel out each other's positive effects. To find a clear pattern between a permutation of code smell refactoring types and energy consumption, experiments with more permutations on a bigger corpus of Android apps are desirable.

> Insight 2: The energy impact over all permutations of code smell refactorings in the selected Android apps was small. However, specific permutations of code smell refactorings should be used with caution as their energy impact might vary strongly depending on the selected Android app.

Form the results related to RQ2 we see that there was no significant impact of code smell refactoring on execution time. In our experiment, we neither observed significant increase nor decrease of execution time due to code smell refactoring in the selected Android apps. This is in contradictory to the results reported by Verdecchia et al. [14]. They reported that observed energy reduction was due to performance-related improvements. Therefore, more experimental evidence is required to confirm the assumption that a trade-off between execution time and energy consumption exists.

> Insight 3: Significant reduction in energy consumption of Android apps does not necessarily correlate with significant reduction or increase of execution time.

We characterised the selected Android apps in section III. D. The ages of the apps 'Todo-List' and 'Openflood' were approximately the same. However, there was a significant difference in the number of code smells refactored in these two apps. Also, the number of contributors to both projects were equal. This suggests that for Android apps of the same age and with the same number of contributors, the probability of detected code smells could be dependent on the size of the projects and also possibly on the experience and knowledge of the contributors. The 'Todo-list' app was considerably bigger in size than the 'Openflood' app and the number of commits in the 'Todo-list' app was much higher than the number of commits in the 'Openflood' app. However, it is worth noting that the code of the 'Openflood' app was written with maintainability in mind.

> Insight 4: We can only expect to see an impact on energy consumption if there is much to refactor. The probability of detecting code smells in Android apps seems to depend on factors like size of the app, age of the app, experience of contributors to the app, and number of contributors to the app.

The 'Calculator' app was larger than the other two apps in all the above stated attributes, i.e., size, age, number of commits, and number of contributors. This might explain that the impact of refactoring on the energy consumption could become very high both when refactoring only single code smell types and also when refactoring all five code smell types together. However, it remains unclear why the refactoring of code smells of certain permutations of code smell types can increase the energy consumption by up to 7.2%, while the refactoring of code smells of a single code smell type generally decreases energy consumption with the only exception of code smell type 'God Class' which shows a small increase of 1%.

## VI. THREATS TO VALIDITY

In this section, we discuss the possible threats to validity of our experiment. The apps chosen for experimentation were selected from F-Droid but to ensure that they were representative of a real world Android apps we checked that these apps are also available on Google play store and have more than 10K downloads and a minimum rating of 4 stars.

We used JDeodorant for detecting code smells and for generating candidate refactorings. The accuracy of this tool might affect the accuracy of the results. We selected JDeodorant because it complies with Mens et al. [26] list of activities that should be followed by a good detection and refactoring tool. JDeodorant has been used in various previous studies for code smell detection, based on the above two reasons we think that the results produced by JDeodorant were reliable. However, we cannot ensure that the same candidate code smell refactorings will be produced if this study is replicated using another code smell detection tool.

Power measurements were recorded using the Monsoon power monitor which recorded power consumption at a rate of 5KHz. The same specification has been used in previous studies so sampling frequency is sufficient to produce precise readings. Power measurements were made for the apps plus Android OS related activities. We tried to minimize this threat by using adb logs and matching the timestamps with the energy trace to filter out any unwanted readings. We repeated the tests 10 times for each version and included the averaged reading to further mitigate variations in the readings. However, recording the logs for the app introduces an overhead which in itself could be energy consuming. But as this overhead was constant between the versions of the app it could be ignored.

It is important to note that in order to replicate the results of this experiment the same phone model and Android OS version, measurement tools and app version should be used. The reason is that different mobile phones with different Android OS versions might use different levels of energy. Changing the version of the apps might result in failure of our tests.

We only analyzed three apps, therefore, the results might not be generalizable. Increasing the magnitude of study in terms of a number of included apps in future work is desirable. Increasing the number of apps in the experiment also exponentially increases the number of app versions that need to be tested. As writing test scenarios, checking each candidate refactoring before applying and mapping of Android projects to Eclipse Java projects is a time consuming and slow process that requires manual effort, we limited this experiment to three apps. Nevertheless, we did consider apps from two different app categories of Google play store.

## VII. CONCLUSION AND FUTURE WORK

Previous studies suggested that developers care about code smells and refactoring. However, not many studies discuss the energy impact of code smell refactorings.

The research presented in this paper extends previous studies by investigating the energy impact of refactoring five code smell types (first individually per type and then in permutations) of native Android open source apps. We also studied the impact of using the code smell refactorings on the execution time of native Android open source apps. Our results indicate that the maximum energy reductions recorded are 10.8% and 10.5% for refactoring code smell 'Duplicated code' and 'Type Checking' respectively. Specific permutations of code smell refactorings should be used with caution, as their energy consumption impact might differ strongly between the selected Android apps. We observe neither significant increase nor decrease of the execution time in selected Android apps. Which indicates that reduction in energy consumption due to code smell refactorings and change in execution time are not directly related. The probability of detecting code smells in Android apps seems to depend on factors like size of the app, age of the app, experience of contributors to the app and number of contributors to the app.

Overall, we conclude the following: when refactoring code smells of a single type only then on can expect to achieve medium to large decrease in energy consumption. We observed only few cases where refactoring code smells of a single type resulted in a small increase of energy consumption. Among the three apps that we analyzed this was the case in the app that was the smallest in size and seemed to have the best code quality. In order to generalize this and other results of our study, we plan to conduct a similar study with a larger set of apps.

ACKNOWLEDGMENTS

This work is supported by the institutional research grant IUT20- 55 of the Estonian Research Council and the Estonian Center of Excellence in ICT research (EXCITE).